\documentclass[fleqn,usenatbib]{mnras}

\usepackage{newtxtext,newtxmath}

\usepackage[T1]{fontenc}
\usepackage[table,xcdraw]{xcolor}

\usepackage{graphicx}   
\usepackage{color}
\usepackage{todonotes}

\newcommand\highlight[1]{#1}

\title[A Digital Correlator Upgrade for AMI]{A Digital Correlator Upgrade for the Arcminute MicroKelvin Imager}

\author[J. Hickish et al.]{
Jack Hickish,$^{1,2}$\thanks{E-mail: jackh@berkeley.edu}
Nima Razavi-Ghods,$^{2}$
Yvette C. Perrott,$^{2}$
David J. Titterington,$^{2}$
\newauthor
Steve H. Carey,$^{2}$
Paul F. Scott,$^{2}$
Keith J. B. Grainge,$^{3}$
Anna M. M. Scaife,$^{3}$
\newauthor
Paul Alexander,$^{2}$
Richard D. E. Saunders,$^{2,4}$
Mike Crofts,$^{2}$
Kamran Javid,$^{2}$
\newauthor
Clare Rumsey,$^{2}$
Terry Z. Jin,$^{2}$
John A. Ely, $^{2}$
Clive Shaw,$^{2}$
Ian G. Northrop,$^{2}$
\newauthor
Guy Pooley,$^{2}$
Robert D'Alessandro,$^{2}$
Peter Doherty,$^{2}$
Greg P. Willatt.$^{2}$
\\
$^{1}$Radio Astronomy Laboratory, University of California, Berkeley, CA 94720, USA\\
$^{2}$Astrophysics Group, Cavendish Laboratory, 19 J. J. Thomson Avenue, Cambridge CB3 0HE\\
$^{3}$Jodrell Bank Centre for Astrophysics, Alan Turing Building, School of Physics and Astronomy, University of Manchester, \\
Oxford Road, Manchester, M13 9PL\\
$^{4}$The Kavli Institute for Cosmology, Cambridge, CB3 0HA\\
}

\date{Accepted XXX. Received YYY; in original form ZZZ}

\pubyear{2017}

\begin{document}

\label{firstpage}
\pagerange{\pageref{firstpage}--\pageref{lastpage}}
\maketitle

\begin{abstract}
The Arcminute Microkelvin Imager (AMI) telescopes located at the Mullard Radio Astronomy Observatory near Cambridge have been significantly enhanced by the implementation of a new digital correlator with 1.2\,MHz spectral resolution. This system has replaced a 750-MHz resolution analogue lag-based correlator, and was designed to mitigate the effects of radio frequency interference, particularly that from geostationary satellites \highlight{which are visible from the AMI site when observing at low declinations}. The upgraded instrument consists of 18 ROACH2 Field Programmable Gate Array platforms used to implement a pair of real-time FX correlators -- one for each of AMI's two arrays.  The new system separates the down-converted RF baseband signal from each AMI receiver into two \highlight{sub-bands, each of which are filtered to a width of 2.3\,GHz and digitized at 5-Gsps with 8 bits of precision.} These digital data streams are filtered into 2048 frequency channels and cross-correlated using FPGA hardware, with a commercial 10\,Gb Ethernet switch providing high-speed data interconnect. Images formed using data from the new digital correlator show over an order of magnitude improvement in dynamic range over the previous system. The ability to observe at low declinations has also been significantly improved.
\end{abstract}




\section{Introduction}
The Arcminute Microkelvin Imager (AMI) instrument consists of two synthesis radio telescopes located at the Mullard Radio Astronomy Observatory near Cambridge. AMI, which operates in the 12--18\,GHz frequency band, was designed primarily for the study of galaxy clusters by observing secondary anisotropies in the cosmic microwave background (CMB) arising from the Sunyaev-Zel'dovich (SZ) effect \citep{sz}.

AMI is made up of two arrays: the Small Array (SA), and the Large Array (LA). The SA telescope comprises ten 3.7\,m diameter paraboloid dishes in a compact configuration
and is designed for observing structures on arcminute scales. The LA, which was created by reconfiguring the eight 12.8\,m dishes of the Ryle Telescope
has an angular resolution of 0.5 arcminutes and has approximately ten times the flux-density sensitivity of the SA. The LA, observing concurrently with the SA, is used to measure the intensities of contaminating small-diameter radio sources.

The original AMI correlator \citep{Zwart2008} was of an analogue "XF" design, whereby the antenna signals were cross-correlated over a range of delays and then Fourier transformed into spectral channels.  The original correlator recorded visibilities in eight frequency channels over AMI's 12--18\,GHz band, providing limited capability for recognising and removing interfering signals as well as limited ability to mitigate chromatic aberration out to the edge of field of view. As a result, the sensitivity of the instrument was significantly reduced, particularly at low declinations, where interference from geostationary satellites could result in up to 90\% of the data being unusable.  The XF system had a performance that was also limited by the path-length inaccuracies of analogue delay compensation and analogue correlation, which resulted in direction-dependent systematic errors and poor dynamic range.

Starting in 2014, a project was undertaken to build a digital correlator for the telescope based on the second-generation Reconfigurable Open Architecture Computing Hardware board (ROACH2\footnote{\url{https://casper.berkeley.edu/wiki/ROACH-2_Revision_2}}), developed by the Collaboration for Astronomy Signal Processing and Electronics Research (CASPER\footnote{\url{https://casper.berkeley.edu}}) and used extensively in radio-astronomy digital signal processing (DSP) applications \citep{hickish-casper}. ROACH2 is a powerful real-time signal processing platform based around a Xilinx Virtex 6 Field Programmable Gate Array (FPGA) which provides up to 80\,Gbps of \highlight{Ethernet-based} I/O, and may be interfaced with a variety of analogue to digital converters (ADCs).
The primary goal of the project was to equip AMI with a digital correlator providing over one hundred times the frequency resolution of the original system with superior inter-channel rejection, allowing better frequency-confinement of interfering signals as well as overcoming some of the limitations of analogue lag correlator.
The wide bandwidth of the AMI receiver required an optimised FPGA correlator design, capable of operating at the relatively high clock speed of 312.5\,MHz, to be developed. This correlator, which \highlight{comprises an independent instrument for each of the AMI arrays uses 18 ROACH2 boards (ten for the SA, and eight for the LA) to digitise, channelise and then cross-correlate analogue inputs from the antennas in the AMI arrays.}  Communication between boards in each AMI array is achieved using an industry-standard 10\,Gb Ethernet (10\,GbE) network, with an off-the-shelf switch providing interconnect. This so-called "Packetized Correlator" architecture was pioneered by the CASPER group \citep{Parsons2008, hickish-casper} and has been well proven at multiple radio-telescopes \citep{kocz-leda, Foley01082016, eovsa, swarm}.


%

In this paper we summarize the design of, and first commissioning results from the new AMI correlator. In Section~\ref{sec:arch-overview} we describe the architecture of the system deployed, which comprises an overhauled analogue front-end and a new digital back-end. In Section~\ref{sec:dig-implementation} details of the implementation of the digital back-end are presented. The instrument's control system and data reduction pipeline are outlined in Sections~\ref{sec:control} and \ref{sec:reduction}, respectively, with commissioning results from the new system given in Section~\ref{sec:results}.

\section{Architecture Overview}\label{sec:arch-overview}

\highlight{The AMI correlator upgrade continues to use the front-end systems developed in \cite{Zwart2008}, with a new final IF stage and digital signal processing backend. A summary of the telescope specifications is given in Table~\ref{table:TECHDATA}.}

\begin{table}
\centering
\caption{Summary of AMI technical data. Specifications pertaining to frequency coverage (slightly reduced) and frequency resolution (dramatically increased), which have changed since \citet{Zwart2008} appear in bold.} 
\label{table:TECHDATA}
\begin{tabular}{c|cc}
           & Small Array   & Large Array \\
\hline
\hline
Antenna diameter (m)   & 3.7 & 12.8 \\
Antenna efficiency  & 0.75 & 0.67 \\
Number of antennas  & 10 & 8 \\
Number of baselines  & 45 & 28 \\
Baseline lengths (m)  & 5--20 & 18--110 \\
Primary beam ($@$15.5 GHz)  & 20\arcmin{.} & 5\arcmin{.} \\
Synthesized beam  & $\approx$ 3\arcmin & $\approx$ 30\arcsec \\
Flux sensitivity (mJy s$^{1/2}$) & 30  & 3 \\
Declination range ($^{\circ}$)  & > -15 & > -20 \\
Elevation limit ($^{\circ}$)  & +20 & +5 \\
\textbf{Observing frequency (GHz)}  & \multicolumn{2}{c}{13.1--17.9}  \\
\textbf{Number of channels}  & \multicolumn{2}{c}{4096}  \\
\textbf{Channel bandwidth (MHz)}  & \multicolumn{2}{c}{1.22}  \\
System temperature (K)  & \multicolumn{2}{c}{25}  \\
Polarisation measurement  & \multicolumn{2}{c}{I+Q}  \\
\end{tabular}
\end{table}

\subsection{Analogue Front-End}

\begin{figure}
 \centering
 \includegraphics[width=\columnwidth]{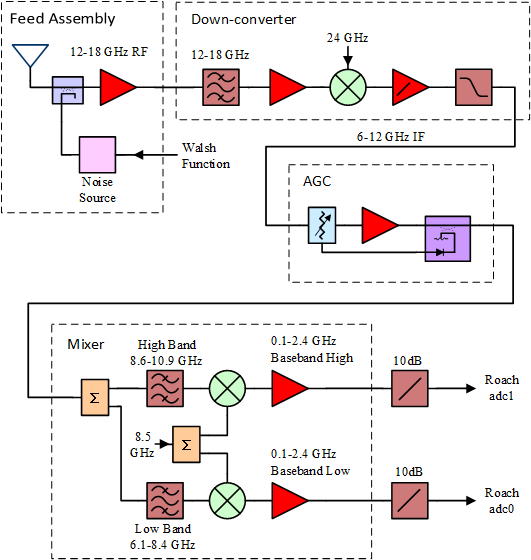}
 \caption{The AMI RF and IF system diagram. Isolators and attenuators used in the chain are not shown. The \emph{Mixer} units in the system are newly developed as part of the correlator upgrade.}
 \label{fig:AMIRF}
\end{figure}

The AMI front-end system consists of a feed assembly, first down-converting mixer  and second mixer stage as shown in Figure~\ref{fig:AMIRF}.  Each feed assembly contains the horn antenna, cryostat, radio-frequency (RF) amplifiers, noise injection, bias box and the cryogenic support system. Each cryostat has three temperature zones separated by thermal shielding; 15--20\,K, 50--70\,K and 300\,K. There are two RF amplifiers: the first is mounted on the 20\,K coldhead of the cryostat, and the second is at 300\,K. Noise at a constant level may be switched into the front-end via a noise source which is outside the cryostat on the SA, and inside the cryostat at 300\,K on the LA.  A detailed description of the \highlight{feed assembly, down-converter, and Automatic Gain Control (AGC) systems, which have not been modified as part of this work,} is given in \cite{Zwart2008}.

The RF (12--18\,GHz) signals from the antenna feeds are down-converted to an intermediate-frequency (IF) band of 6--12\,GHz using a 24\,GHz local oscillator (LO). Phase-switching is provided by modulating this LO signal by Walsh functions which are generated by the digital correlator. The IF is fed from the antenna hub to a correlator room where an AGC unit is used to maintain a constant power level prior to further down-mixing. 

A second mixer assembly splits the down-converted IF output into two 2.3\,GHz wide sub-bands by first filtering the low and high bands and then mixing with an 8.5\,GHz common local oscillator. Custom-manufactured low pass filters are used to further limit the bandpass of the baseband signal prior to being fed into ADCs. The terms "low band" and "high band" used here to describe the two AMI correlator basebands refer to the frequencies of the mixer bandpass filters, not their respective RF input frequencies, as shown in Table \ref{table:AMIBANDS}.  The second mixer assembly for the new system provides approximately 5\,GHz of the original 6\,GHz AMI frequency channel.  The improvements in dynamic range are evident even with the loss in sensitivity due to processing less channel bandwidth.

\begin{table}
\centering
\caption{The AMI frequency bands.} 
\label{table:AMIBANDS}
\begin{tabular}{c|ccc}
           & RF (GHz)   & IF (GHz)  & Baseband (GHz) \\
\hline
\hline
Low Band   & 15.6--17.9 & 6.1--8.4  & 0.1--2.4 \\
High Band  & 13.1--15.4 & 8.6--10.9 & 0.1--2.4 \\
\end{tabular}
\end{table}

\subsection{Digital Back-End}
The AMI digital correlator is implemented using the standard "FX" architecture favoured by most modern correlators. In this architecture analogue signals from each antenna are first digitised before being split into spectral channels using an FFT-based algorithm (the "F" processing stage). Spectra from pairs of antennas are then multiplied and time-averaged on a per-frequency-channel basis (the "X" processing stage). The F-stage processing of such a correlator can be trivially parallelised over the multiple antennas in an array, whilst the X-stage computation is parallel over multiple frequency channels. Interconnect exists between the F and X processors to facilitate a data transpose---often referred to as a \emph{corner-turn}---which allows data to be aggregated appropriately so as to allow parallel processing.

The AMI correlator uses FPGAs to perform the F and X processing, owing to their support for high input and output data rates and ease of interfacing to high-speed ADCs. Interconnect is provided by commercial off-the-shelf (COTS) 10\,Gb/s Ethernet (10\,GbE) switches, which offer cost-effective, flexible and reliable performance. Such switches also support useful functionality such as multicast (point to multi-point data transmission) which may be used in future correlator operating modes. This ``packetised correlator'' architecture has been well-developed by the CASPER collaboration, which provides and maintains software and firmware libraries to facilitate rapid deployment of digital astronomical systems.

\begin{figure}
 \centering
 \includegraphics[width=0.6\columnwidth]{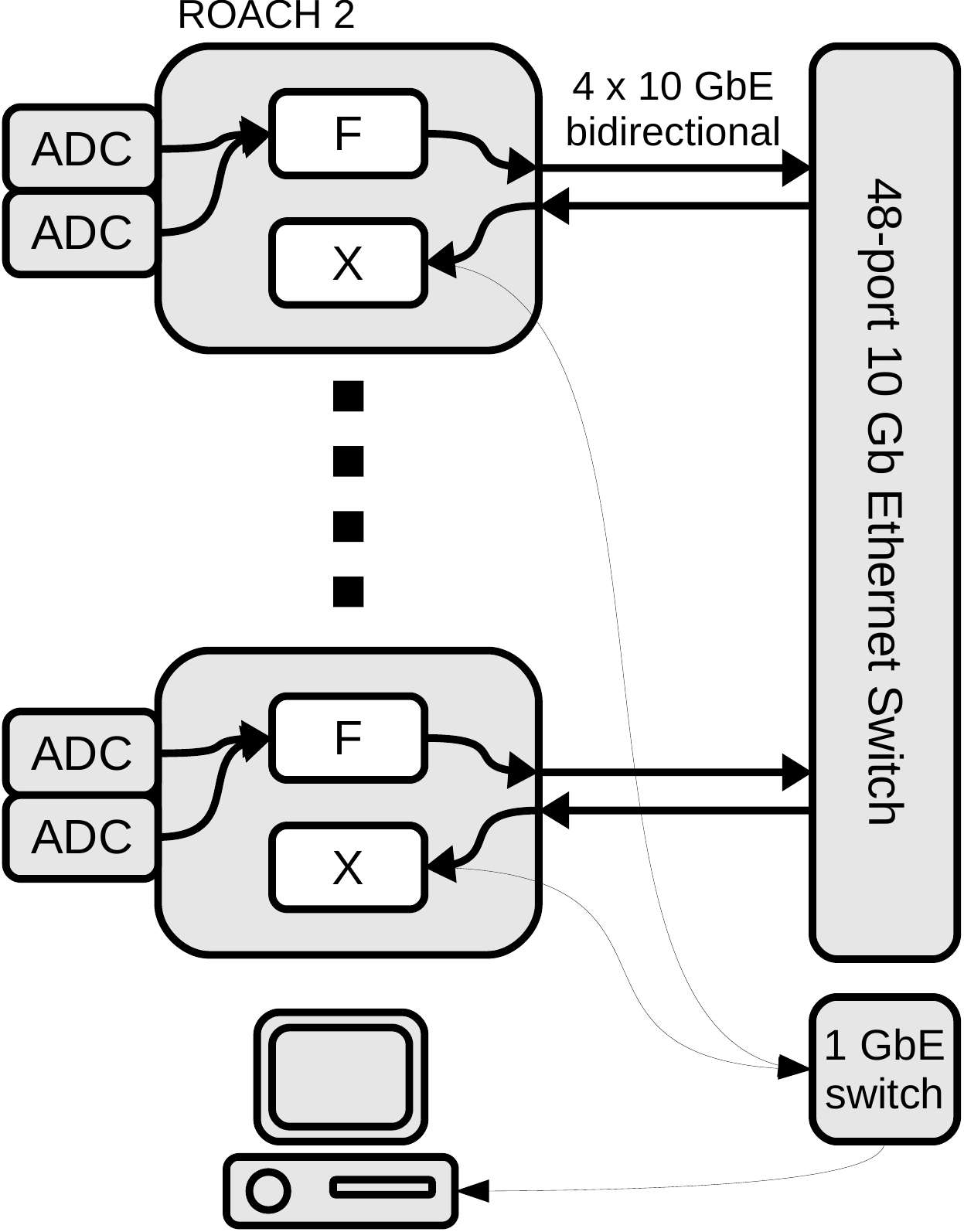}
 \caption{The top-level correlator architecture. Digitisation is performed by a pair of 5\,Gsps ADC cards interfaced to a ROACH2. The ROACH2 firmware implements both channelisation and cross-multiplication functionality, with a 10\,GbE switch providing interconnect between the two phases of processing. Averaged correlation results are output over a separate 1\,Gb Ethernet network.}
 \label{fig:top-block}
\end{figure}

Though not required by the packetised architecture, the AMI correlator design uses the same physical processors for its F and X stages (Figure~\ref{fig:top-block}). This allows the complete system to be efficiently implemented with minimal hardware. Once correlation matrices have been computed and averaged for an appropriate time window, these are sent over a 1\,Gb Ethernet network where the results are aggregated by a standard x86 GNU/Linux server.

\subsubsection{Digital Hardware}
The digital processing platform chosen for the AMI correlator is the mature and ubiquitous CASPER-designed ROACH2 board\footnote{\url{https://casper.berkeley.edu/wiki/ROACH-2_Revision_2}}. This is a general-purpose FPGA platform, built around a Xilinx Virtex 6 (XC6VSX475T-1FFG1759C) FPGA, supported by four 72\,Mb QDR memory chips, 1\,GB of DRAM, and up to eight 10\,GbE interfaces via a pair of SFP+ mezzanine cards. The ROACH2 also provides two 40-pair LVDS interfaces via Z-DOK connectors, which can be used to interface the ROACH2's FPGA to a variety of CASPER-supported ADC cards.

For the AMI correlator, both Z-DOKs are populated with CASPER ADC1x5000\footnote{\url{https://casper.berkeley.edu/wiki/ADC1x5000-8}} cards. Each of these hosts an e2v EV8AQ160: a quad-core digital sampler capable of sampling a single RF input at up to 5\,Gsps. The ADC1x5000 was designed by the Academia Sinica Institute of Astronomy and Astrophysics (ASIAA) and has been extensively characterised by \cite{Patel2014} as part of the development of a new wideband correlator for the Submillimetre Array \citep[SMA,][]{swarm}. 

Each ROACH2 in the AMI system is fed with two timing signals derived from a COTS Trimble Thunderbolt E GPS-disciplined oscillator, which provides a pulse-per-second (PPS) reference and 10\,MHz frequency standard. The PPS is distributed to each ROACH2 in the system via a 16-way buffered splitter and is used to synchronise and timestamp the data outputs from each board. The 10\,MHz reference is used to derive a 2500\,MHz clock, which is amplified and split so that it may drive the 5000\,Msps ADC samplers synchronously.

\begin{figure}
 \centering
 \includegraphics[width=\columnwidth]{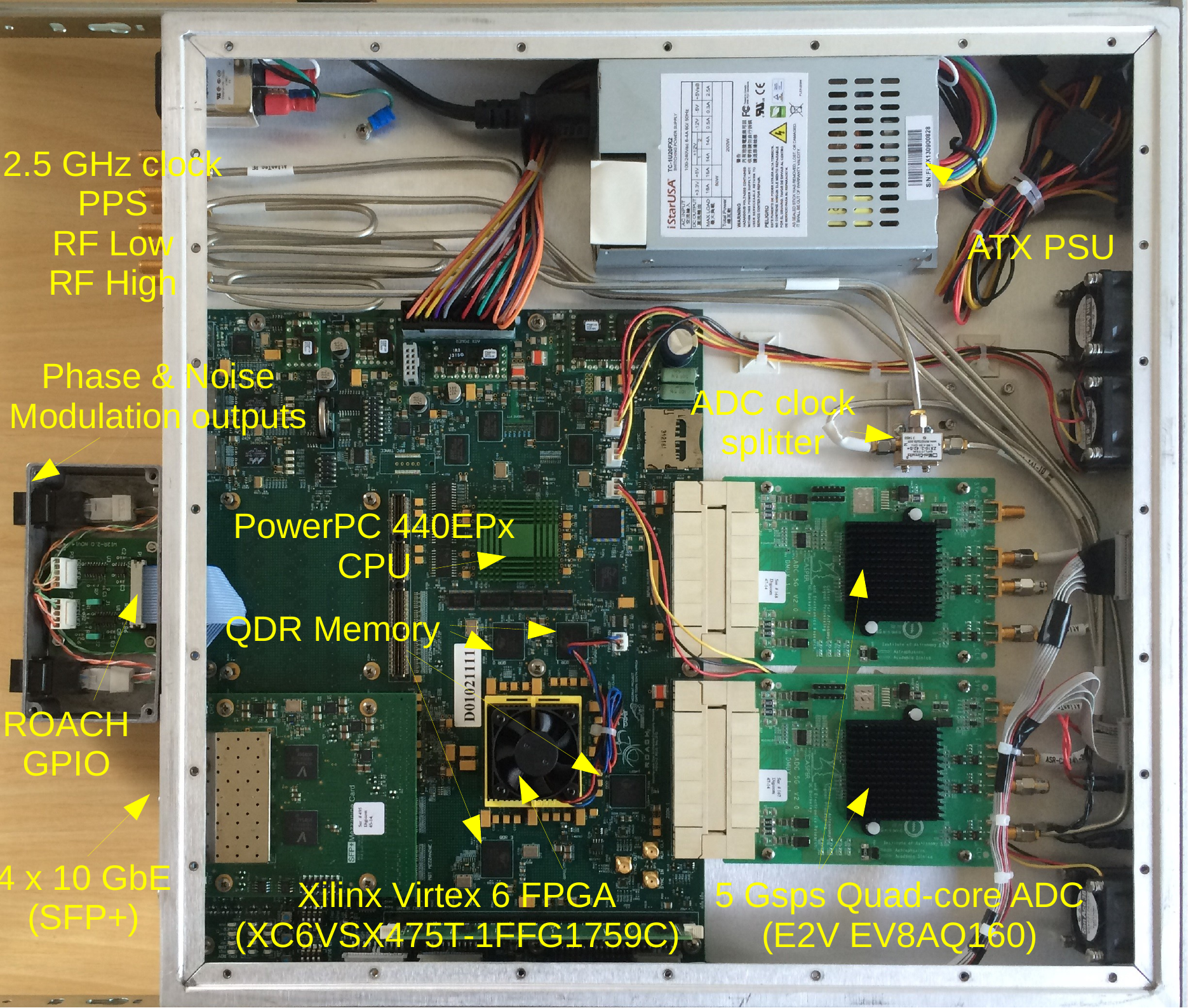}
 \caption{The CASPER ROACH2 platform on which the AMI digital correlator is implemented. Here the board is shown with two 5\,Gsps ADC daughter cards, a quad-SFP+ mezzanine card, and a Walsh switch interface box.}
 \label{fig:ROACH2}
\end{figure}

The complete ROACH2 processing node is shown in Figure \ref{fig:ROACH2}. The node interfaces comprise:
\begin{enumerate}
 \item An SMA input a for 2.5\,GHz clock (used to derive the 5\,Gsps ADC sampling rate).
 \item An SMA input for a TTL Pulse-per-second (PPS) time reference.
 \item SMA inputs for the high and low baseband inputs from the analogue front-end of a single AMI antenna.
 \item Level-shifting circuitry to allow interfacing of the ROACH2 FPGA with the AMI phase and noise modulation infrastructure (see Section \ref{sec:walsh}).
 \item A 1000BASE-T Ethernet interface to the FPGA, for data output.
 \item A 1000BASE-T Ethernet interface to ROACH2's on-board CPU, which is used for controlling and monitoring the board.
 \item Four SFP+ connectors, each providing a 10\,GbE interface.
\end{enumerate}

A key part of the correlator hardware infrastructure is the 10\,GbE switch facilitating interconnection between the 10 (8) boards in the SA (LA) system. This is a Mellanox SX1012, 12-port, 40\,Gb Ethernet switch, capable of operating as a 48-port 10\,GbE switch using interconnecting cables to connect each of the switch's QSFP+ ports to four independent SFP+ interfaces. 

\begin{figure}
 \centering
 \includegraphics[width=\columnwidth]{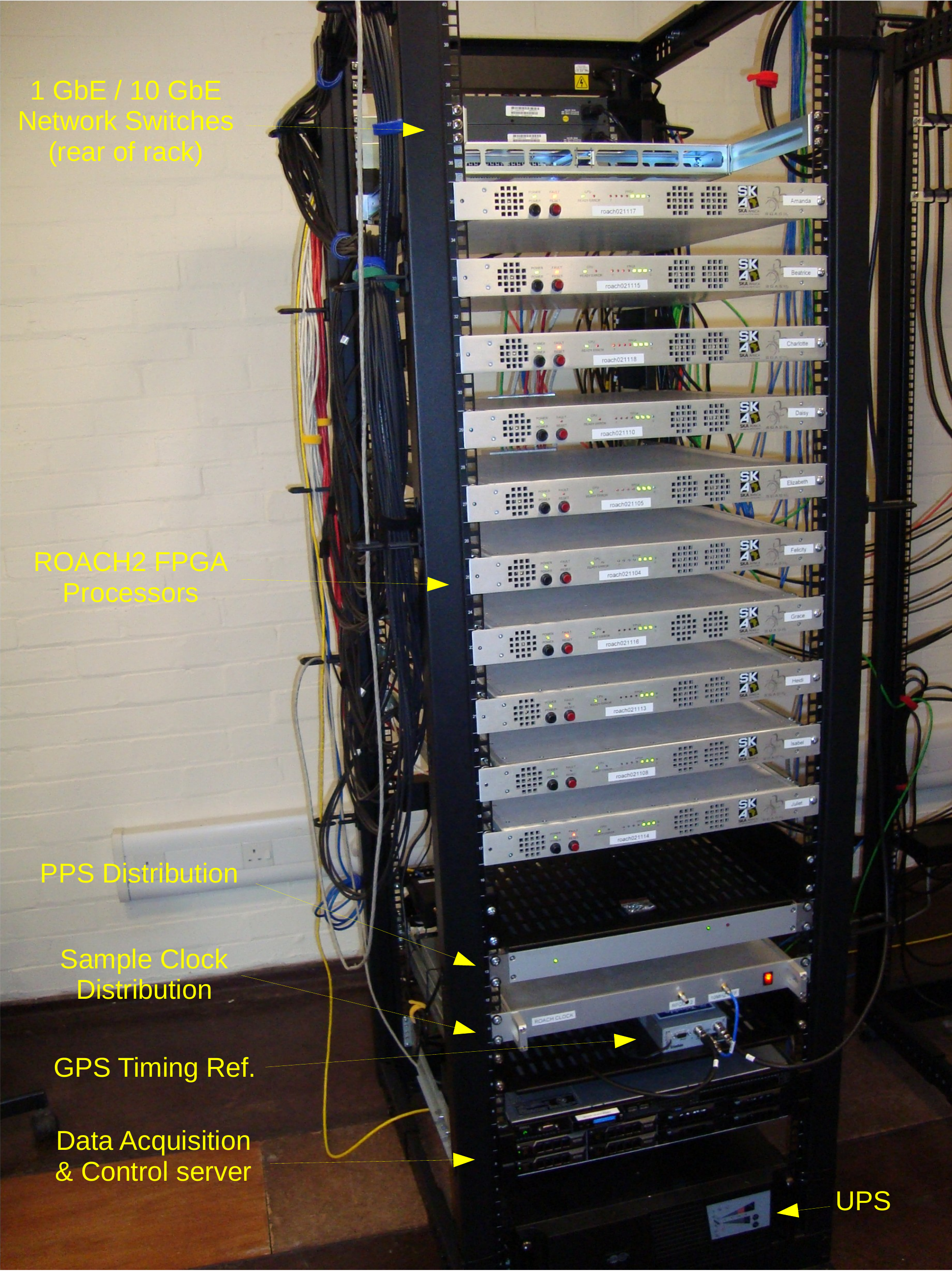}
 \caption{The digital correlator rack for the 10-antenna AMI Small Array.}
 \label{fig:digital-rack}
\end{figure}

The complete rack of digital equipment for the SA correlator is shown in Figure~\ref{fig:digital-rack}.

\section{Correlator Firmware}\label{sec:dig-implementation}

\subsection{Analogue-to-Digital Converters}
The CASPER collaboration (and in particular CASPER members in the Submillimetre Array correlator group) provide open-source interface firmware\footnote{\url{https://github.com/casper-astro/mlib_devel}} and software\footnote{\url{https://github.com/sma-wideband/adc_tests}} to stream data from the e2v EV8AQ160 ADC into the ROACH2 board via its Z-DOK connector.
This interface expects data to be presented to the FPGA at a quarter of the ADC sample rate, over four parallel 8-bit buses. At a maximum sampling rate of 5\,Gsps, this results in 32 parallel data lines, each running at 1.25\,Gbps. Once captured by the FPGA, these parallel samples are demultiplexed by a further factor of four, so that on every FPGA clock cycle, 16~samples are processed in parallel. In the AMI design the signal processing pipeline is clocked synchronously with the ADC, at a rate of 312.5\,MHz. At the time of design, the interface provided by CASPER did not function at the full 5\,Gsps rate of which the ADCs are capable. As part of this work, the interface has been modified to increase performance and released back to the CASPER community.
\highlight{We note that since the ADC used by AMI is of multicore design, data are liable to artifacts caused by mismatch in the timing, gain, and voltage offsets of the different sampler cores \cite{Patel2014}. Currently the AMI system flags the frequency channels associated with artifacts from voltage offsets (this is the channel at the center of the digitized band), but does not correct for timing and gain mismatches. Further work is needed to assess the effect of these mismatches on the broadband images AMI produces, but they are not thought to limit the performance of the array, due to the suppression effects provided by fringe-tracking, and time- and frequency-averaging.}

\subsection{Front-end noise injection \& phase switching control}
\label{sec:walsh}
In the original AMI analogue correlator, phase and noise modulation functions were generated by a bank of look-up tables driven by a 65536\,Hz counter. In order to simplify the new digital system, it was decided that the correlator FPGAs should be responsible for generation of the switching signals. In this way the switching frequencies are synchronous with the ADC sampling, and can easily be chosen such that an integral number of switching periods occur in each channelisation window and correlator integration period.

In order that the FPGAs are able to drive the existing front-end switching infrastructure, simple interface boards
were constructed to convert a pair of 1.5\,V ROACH2 general-purpose IO (GPIO) outputs to drive 5\,V differential signals over Category 5 UTP cable. Though unused in the AMI correlator, the interface boards are also capable of converting a pair of differential signals to single-ended 1.5\,V inputs, which may be used to drive ROACH2 GPIO pins.


\subsection{F-Engine}
\label{sec:f-engine}
\begin{figure*}
 \centering
 \includegraphics[width=2\columnwidth]{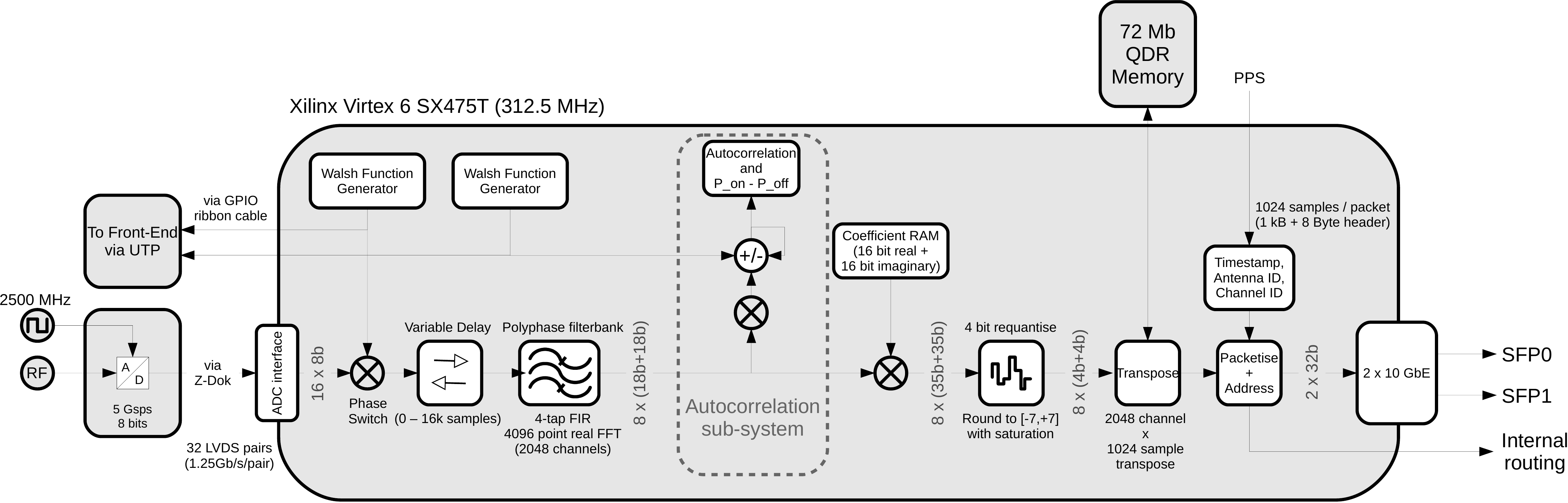}
 \caption{A schematic depiction of the AMI F-Engine firmware. Two such F-Engines are instantiated on each ROACH2 FPGA.}
 \label{fig:ami-f}
\end{figure*}

The bulk of the new AMI correlator design comprises the ``F-Engine'' processing pipeline; the channelisation of pairs of 5\,Gsps data streams into 2048 critically-sampled subbands, which are output over 10\,GbE as User Datagram Protocol (UDP) data streams (Figure~\ref{fig:ami-f}).
Stages of the pipeline are:

\paragraph*{Phase-demodulation} Immediately after samples are captured into the FPGA, the phase modulation applied in the antenna's first mixer stage is removed by a simple multiplication of ADC samples by $\pm 1$. The modulation/demodulation pattern used is unique to each antenna, and stored in the firmware in a runtime-programmable look up table. A programmable delay between the modulation GPIO output (Section~\ref{sec:walsh}) and the internal demodulation signal allows compensation for delays associated with cable lengths of the control and RF signals.

\paragraph*{Coarse delay} After demodulation, each antenna's digital data stream may be delayed by up to 16,384 ADC samples. This delay allows for compensation of geometric delays in the array (the largest baseline in the LA is approximately 110\,m) and RF cabling. The desired delays are calculated by the telescope control computer based on the current pointing of the array, and any pre-computed delay calibration solutions. The frequency resolution of the AMI digital correlator is $1.22$\,MHz, giving an inverse-channel-bandwidth of around 800\,ns. This inverse-bandwidth sets the accuracy with which variable delays must be applied to avoid losing coherence of the antenna signals; for AMI, it is sufficient to update coarse delays on $\sim$second timescales. For simplicity of data analysis, the correlator control software ensures that delay updates are applied synchronously with new visibility accumulations, such that no coarse delay changes occur mid-integration.

\paragraph*{Polyphase Filterbank} The largest component of the F-Engine processing pipeline is a polyphase filterbank (originally proposed by \cite{bellanger}, see \cite{harris-haines, price2016spectrometers} for relevant overviews), which breaks the 5\,Gsps data streams into 2048 critically sampled subbands, each of bandwidth 1.22\,MHz. The CASPER group provide parameterised finite impulse response (FIR) filter and fast Fourier transform (FFT) libraries for implementing polyphase filterbanks -- these have been utilised here, after various optimisations were applied to reduce FPGA resource utilisation and to maximise clock frequency. \highlight{These optimisations are publically available\footnote{See the \texttt{ami\_devel} branch of the CASPER libraries at \url{https://github.com/jack-h/mlib_devel/tree/ami-devel}} and include:
\begin{itemize}
    \item Reductions in overall FPGA fabric use by reducing control logic.
    \item Improved timing performance of reordering blocks when per-clock enabling is not necessary.
    \item Utilization of FPGA single instruction, multiple data (SIMD) instructions for complex addition and subtraction operations.
    \item Improved timing performance or integer rounding operations by allowing the use of dedicated FPGA arithmetic cells.
\end{itemize}
}

Signals within the channeliser are processed with 18 bits of precision in both channeliser coefficients and data path, resulting in output subbands (spectra) which are complex-valued, with each of the real and imaginary components represented as signed 18-bit numbers.

\paragraph*{Autocorrelation sub-system} After channelisation, the power of each spectrum is computed and fed into a vector accumulator. This provides (without requiring the downstream correlator infrastructure) an averaged power-spectrum for each antenna signal. The autocorrelation subsystem can also demodulate by the Walsh pattern used to drive the noise injection at each of the antenna feeds, providing a measurement of the system-temperature of each antenna. Since the system temperature of antennas varies with weather, constant measurements of these "rain-gauge" values are performed and stored with final data products.

\paragraph*{Requantization} In order to reduce FPGA output bandwidth, data samples are rounded to 4-bit values -- real and imaginary parts in the range $(-7,+7)$ -- with saturation logic prior to being streamed over Ethernet. In order to use the available 4-bits of range most efficiently, before quantisation each of the 2048 subbands associated with an ADC data stream are equalised by multiplying by per-antenna, per-subband coefficients. These coefficients are runtime-programmable, and can be calculated either by examining data from the autocorrelation sub-system or the final correlator data products.

\paragraph*{Data buffering and transmission} The CASPER cross-correlation module used in the correlator \citep{Parsons2008, HickishThesis} expects data to be presented in windows of 1024 samples from each frequency channel in turn. To achieve this, 1024 spectra, each of 2048 channels, must be re-ordered in memory to gather similar channels into contiguous blocks. This reordering operation requires a buffer of $\sim$MB size. Such memory is not available on the FPGA chip itself, so external 72\,Mb Quad Data Rate (QDR) memory chips (of which four are provided by the ROACH2 platform) are used for this reorder operation.

Once data are suitably ordered, they are transmitted as streams of UDP packets over an Ethernet network. The destination address of each packet is determined by the frequency channel associated with the data in that packet's payload, with the end result that each ROACH2 in the network receives data associated with all antennas in the array, but only certain frequency channels.

\subsection{Data Interconnect}

After 4-bit requantization, the data output rate from each dual-band F-engine is 40\,Gbps. These data are transmitted as a stream of UDP packets, with each packet's payload comprising 1024 time samples from a single frequency channel. When transmitted over Ethernet as a stream of UDP packets, transmission protocol overhead -- Ethernet, Internet Protocol, UDP and application headers -- must also be accommodated. In order to transmit the stream over four 10\,GbE links the AMI system takes advantage of the fact that F- and X-processors are located on the same physical hardware. This allows frequency channels with the same source and destination ROACH2 to be routed internally, bypassing the Ethernet interconnect. Such routing reduces the necessary throughput of the interconnection by 10\% (12.5\%) for the Small (Large) AMI array. Including transmission protocol overhead, the aggregate output data rate from each 10\,GbE interface is 9.4 (9.1)\, Gbps for the Small (Large) array. \highlight{The full 2.5\,GHz digitized bands are output, though in the case of the Small Array only 2040 of the total 2048 frequency channels from each band are processed, in order to equally share channels among the 10 processing nodes in the system.}

\subsection{X-Engine}

\begin{figure*}
 \centering
 \includegraphics[width=2\columnwidth]{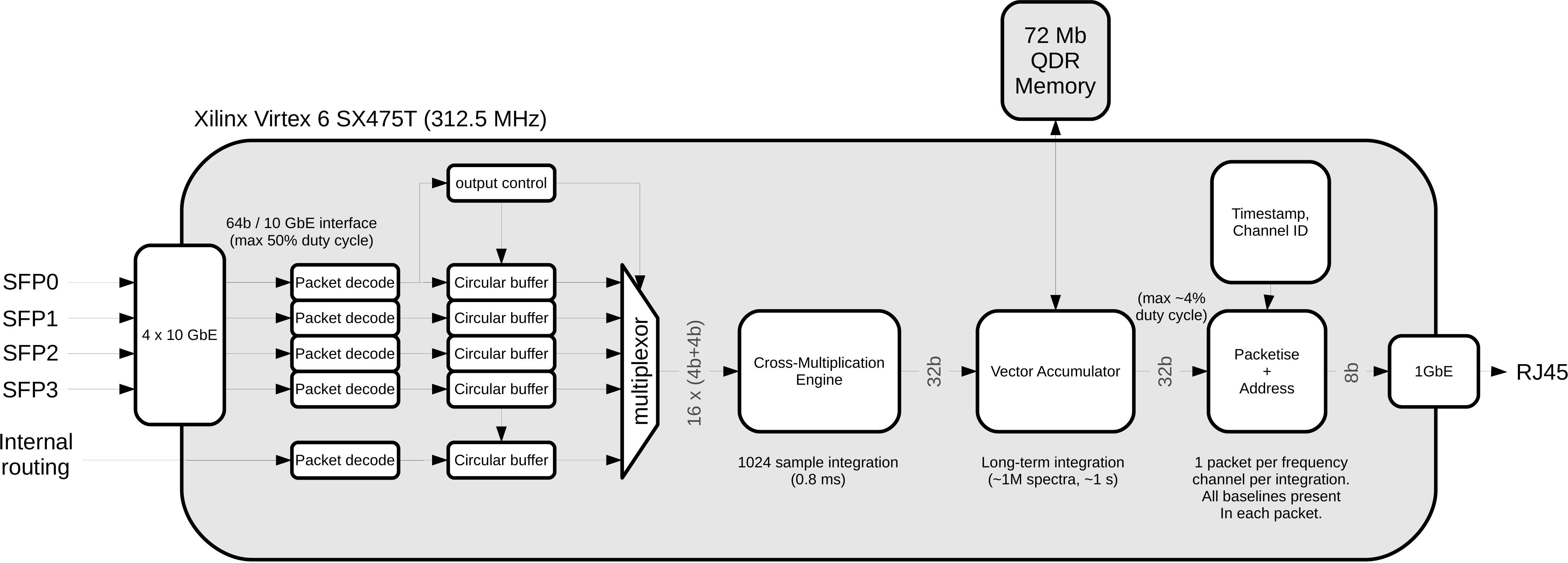}
 \caption{A schematic depiction of the AMI X-Engine firmware.}
 \label{fig:ami-x}
\end{figure*}

The second stage in the correlator is the ``X-Engine'', which is responsible for performing a per-frequency-channel cross-multiplication of data from different antenna pairs. The X-Engine pipeline is shown in Figure~\ref{fig:ami-x}, and its components comprise:

\paragraph*{Input buffering} Data are received from the Ethernet network on the same four SFP+ interfaces used for transmission, as well as an internal routing path. The order that packets from different antenna sources are received from the network is not known in advance -- a circular buffer is used to collect, and appropriately order, data from different antennas prior to cross-multiplication. Once all antenna packets have been received for a given frequency channel, $n$, which is indicated by the arrival of a packet of channel $n+2$, a window of data is streamed into a cross-multiplication engine.

\paragraph*{Cross-Multiplication} The cross-multiplication engine is responsible for taking 1024 samples from a single frequency channel from all antennas, and delivering a visibility matrix integrated over these samples. The correlation engine used in AMI is based on a ``windowed X-Engine'' design by \cite{Parsons2008} which is now maintained by the CASPER community. The AMI version of this module has been ported to Verilog (rather than the CASPER standard of Xilinx System Generator and MATLAB Simulink). \highlight{This makes the module more portable to non-CASPER projects which do not use Simulink and more easily version controlled and simulated using industry-standard tools which are not designed for use with Simulink model files. As part of the porting process, the module was also significantly optimised to minimise FPGA resource utilisation and maximise performance.}

Firstly, AMI's X-engine has a parameterisable input bandwidth, which can be any multiple of the FPGA clock speed. This allows a reduction in control-logic versus multiple instances of a fixed input-bandwidth engine.
Secondly, the cross-multiplication and accumulation cores at the heart of the X-engine have been optimised for 4+4-bit complex inputs, allowing a $75\%$ reduction in multiplier use. This is achieved by offsetting the 4-bit correlator inputs to unsigned values, and appropriately packing pairs of them in 18-bit representations such that four 4-bit multiplications may be computed in a single 18 $\times$ 18-bit operation. A full description of this implementation can be found in \cite{HickishThesis}, which builds on the 4-bit packing speed-ups of \cite{deSouza2007} and is similar to more recent work targeting cross-multiplication of astronomical signals on Graphics Processing Units (GPUs) by \cite{Chime2015}. As a result, the footprint of the cross-multiplication component of the AMI digital correlator has been reduced to a very small fraction of the overall design (Table~\ref{table:resources}).
\highlight{The code for this X-Engine is freely available\footnote{See the \texttt{ami\_devel} branch of the CASPER libraries at \url{https://github.com/jack-h/mlib_devel/tree/ami-devel}} and is provided with a parameterized Simulink wrapper for use in CASPER systems. Though not used in the AMI project, a dual-polarization version of this module is under development, which provides a drop-in replacement for the standard CASPER Simulink module and offers a superset of the CASPER block's parameters.} 

\paragraph*{Vector Accumulator} Per-frequency-channel visibility matrices, which have been integrated over 1024 samples, must be further time-averaged to reduce output data-rate to an acceptable level. This long-term ($\sim$second) averaging uses external Quad Data Rate (QDR) memory for data storage, at the end of which data are transmitted via a 1\,Gb Ethernet interface to a data acquisition and storage server.

\subsection{FPGA Floorplanning}
A significant hurdle in the deployment of the new AMI correlator was achieving timing-closure of the firmware designs at the required 312.5\,MHz FPGA clock rate. Ultimately, this was achieved by constraining the placement of most of the $\sim$1000 multiplier cores and $\sim$400 RAM blocks in the design to small regions of the FPGA chip. \highlight{Placement was manually defined for most modules, but in the case of the FFT cores, a pair of which comprise the majority of the signal processing in the firmware, a set of Python scripts was used to automate constraint generation. These scripts compute the number of multiplier and memory cores needed by each FFT, based on the known number of DSP slices and memory blocks needed by the CASPER FFT butterfly architecture, and the number of butterflies required to implement an FFT meeting the bandwidth and transform length specifications of the AMI channeliser. Constraints are then generated which stripe sequential butterflies over user-defined columns of DSP and memory resources. Having placed these resources, much of the FFT logic, which aims to provide pipelining between arithmetic components, could be removed, resulting in an FFT circuit that is smaller, can be run at higher clock rate, and, critically, has timing performance which is relatively stable when other areas of the FPGA design are modified.}
To the knowledge of the authors, the AMI design represents the fastest clocked ROACH2 in any deployed packetised-correlator system, with the largest instantaneous processed bandwidth per board. We note, however, that the AMI firmware resource requirements (Table~\ref{table:resources}) are modest when compared to some other deployments, such as \cite{swarm}. Experience with newer FPGA platforms also suggests that many of the timing issues encountered during development of the AMI correlator may be unique to the ROACH2's Virtex 6 FPGA. 

\begin{table*}
\centering
\caption{FPGA resource utilisation in the AMI SA correlator design. The AMI LA design uses slightly fewer resources, owing to the smaller number of antennas requiring correlation in the X-Engine stage of the design.}
\begin{tabular}{c|cccc}
           & DSP slices   & BRAMs (36\,kB)   & LUTs & Flip Flops \\
\hline
\hline
Coarse Delay (dual-band)       & 0   & 16   & 2024  & 2314   \\
FIR filter (dual-band)         & 128 & 66   & 3952  & 11890  \\
FFT (dual-band)                & 736 & 112  & 38206 & 100748 \\
X-Engine input buffers         & 0   & 81   & 3162  & 10747  \\ 
X-Engine                       & 98  & 18   & 2002  & 4722   \\

\hline
Total Design             & 1203 & 409  & 105068 & 166051 \\
FPGA Capacity            & 2016 & 1064 & 297600 & 595200 \\
\% Utilisation           & 60   & 38   & 35     & 28     \\
\end{tabular}
\label{table:resources}
\end{table*}


\section{Control System}\label{sec:control}

The control systems for the two arrays are independent, and are described in detail
in \cite{Zwart2008}.  In each case, the only modifications needed for the new correlator have been replacement of the correlator microprocessor by the data acquisition and control server, situated in the digital rack and visible in Figure \ref{fig:digital-rack}, and replacement of the Sun-Microsystems workstation used for telescope and observation control by a Linux desktop system.  Apart from the data acquisition components described below, the rest of the control software, written in C for the Solaris operating system, has simply been recompiled for the Linux platform.

\subsection{Data Acquisition}

The Linux server used for correlator control and data acquisition is a standard single-socket machine, configured with 20\,TB of local disc storage to accommodate several months of raw data.  \highlight{Data are captured from the ROACH2 boards over a 1\,Gbps Ethernet interface (Figure~\ref{fig:top-block}).  A separate 1\,Gbps interface is used for exchanging quick-look data samples and observation metadata with the telescope control system over a local Ethernet connection, as described in Section~\ref{sec:data-handling}}.  This machine also acts as the DHCP/NFS server for the ROACH2 private network.

A suite of Python scripts to configure the ROACH2 boards and perform data collection has been developed for AMI, and is available on github\footnote{\url{https://github.com/jack-h/ami_correlator_sw.git}}.
There are separate scripts for programming and initialising the ROACH2 boards, and for computing the equalisation coefficients for requantization (Section~\ref{sec:f-engine}). 
Acquisition scripts deal with applying coarse path compensation delays to the digital data streams, exchanging metadata and visibilities with the telescope control machine and recording visibility data from the correlator at a programmable dump rate (currently every 0.86\,s) in HDF5\footnote{\url{https://www.hdfgroup.org/HDF5/}} format on the server's local discs.  The arrangement for monitoring the relative system temperature of each antenna using amplitude-modulated injected noise referred to earlier and described in \citet{Zwart2008} is retained, and samples of the demodulated noise, derived from the autocorrelation measurements, are stored with the data for use during calibration processing.

Important features of the software design include the use of a standard configuration file, which allows the same code to be used for both arrays, and a REmote Dictionary Service (Redis\footnote{\url{http://redis.io/}}) key-value data store to share all data among the independently executed processes.  A comprehensive set of utilities for monitoring the state of the correlator and visualising the datasets is also included in the AMI software package.

Independently of data acquisition from the correlator, but at a similar sampling rate, metadata such as antenna pointing and instrumental monitoring information (e.g. cryostat temperatures, AGC readings) are collected by the telescope control machine, and exchanged with the correlator server over the local Ethernet.

\subsection{Data Handling}\label{sec:data-handling}

The HDF5 format data files, one per observation, include correlated visibility data, noise-injection measurements and metadata samples, and accumulate on discs local to the correlator control server.
Typical raw data rates are shown in Table \ref{table:DATARATE}.  These files are the primary data products from the telescope and contain uncalibrated, full spectral resolution data.

\begin{table}
\centering
\caption{Typical AMI data rates.} 
\label{table:DATARATE}
\begin{tabular}{c|cc}
           & Small Array   & Large Array \\
\hline
\hline
Sampling rate (Mbps)   & 16.8 & 11.0 \\
Recording rate (GBph)  & 8 & 5.2 \\
8-hour observation (GB)  & 64 & 42 \\
\end{tabular}
\end{table}


After each observation the HDF5 dataset is converted to FITS-IDI\footnote{\url{http://www.nrao.edu/aips/FITSIDI.pdf}}
format for calibration and further processing using standard reduction packages, and is transferred from the observatory to a data repository at the Cavendish Laboratory.  The conversion script is based on the pyFitsidi\footnote{\url{http://telegraphic.github.io/pyfitsidi}} Python module, and also performs the following operations:
\begin{itemize}
\item \highlight{removal of spike artefacts in a few fixed channels of the ADC readout which are contaminated by ADC core offset mismatches, by flagging these channels to zero.}
\item removal of the channel readout delay phase shift.
\item re-ordering of the ADC channels into increasing RF order.
\item fringe rotation to remove the astronomical path difference over each sub-band.
\item \highlight{amplitude correction to compensate for loss-of-signal due to averaging over 0.86s time windows; this is a function of the fringe rate and
can be as much as a few \% for some of the Large Array baselines.}
\item amplitude calibration for system temperature using the noise-injection system.
\item flagging for instrumental problems (e.g. AGCs, cryostats, pointing, shadowing).
\item optional binning of frequency channels, to reduce the dataset size.
\end{itemize}

In parallel to the above, the raw data samples received from the correlator are binned to 8 $\times$ 0.625 GHz frequency channels by a new task added to the real-time software on the telescope control system, and are recorded in legacy format for quick-look using the in-house software tool, REDUCE.  These 8 channels are chosen to correspond to those used by the old analogue correlator, and availability of these datasets has proved invaluable for data evaluation and diagnostics throughout the development process and during routine operation.

\section{Data Reduction}\label{sec:reduction}

The FITS-IDI files contain visibility data to which instrumental flagging and some initial calibration steps have been applied, and these can now be imported to the standard reduction packages for radio interferometric data such as AIPS\footnote{\url{http://www.aips.nrao.edu/}} and
CASA\footnote{\url{http://casa.nrao.edu}} for full calibration, radio-frequency interference (RFI) flagging and map-making.
A generalised pipeline script has been developed to process these datasets within the CASA framework, and deals with:
\begin{itemize}
\item RFI flagging, both broad-band and at full frequency channel resolution.
\item bandpass (amplitude and phase) and flux calibration, using observations of standard calibrators.
\item phase calibration, using data from interleaved visits to nearby high flux density, unresolved radio sources.
\item imaging.
\end{itemize}

Primary beam correction and mosaicing is currently done in AIPS due to difficulties in importing new primary beam models into CASA.

This is the recommended route for processing AMI data for science.  The 8-channel legacy format datasets described above are also available and can be processed using the in-house REDUCE software to provide useful results for some projects, but lose the benefits of the full spectral resolution that the new correlator provides.

\section{Commissioning Results}\label{sec:results}

\subsection{RFI rejection}

\begin{figure*}
 \centering
 \includegraphics[trim={1.5cm, 1.5cm, 0.5cm, 0.8cm}, clip=,width=\columnwidth]{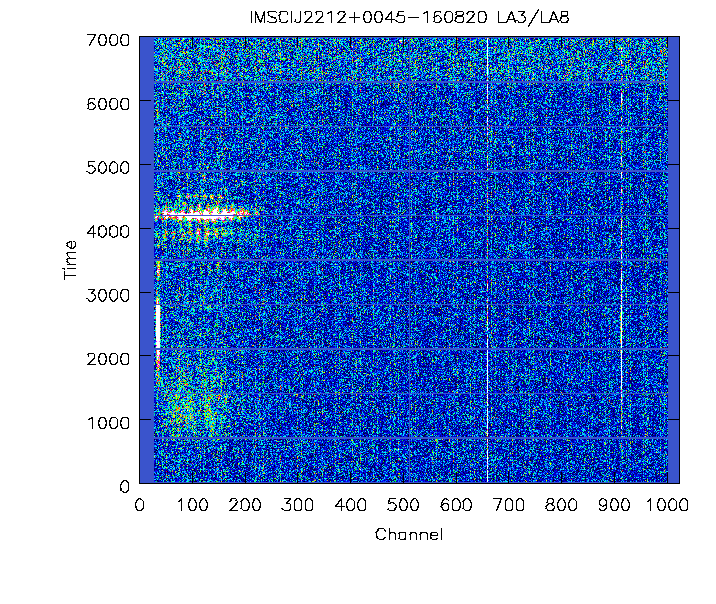}
 \includegraphics[trim={1.5cm, 1.5cm, 0.5cm, 0.8cm}, clip=,width=\columnwidth]{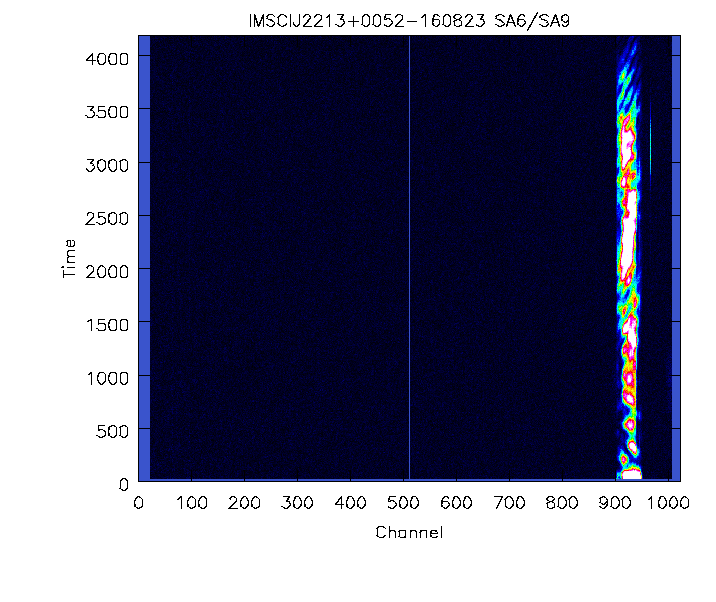}
 \caption{`Waterfall' plots of a low-declination ($\approx\,0^{\circ}$) field observed with the LA (SA) on the left (right), showing time against channel for the lower (upper) half of the bandwidth on a single baseline.  The RFI can be clearly seen as bright signals localised in frequency and time and is easily removed using automated algorithms such as `rflag' in CASA; see Fig~\ref{Fi:RFI_flagging}.}
 \label{Fi:RFI}
\end{figure*}

\begin{figure*}
 \centering
 \includegraphics[trim={1.5cm, 1.5cm, 0.5cm, 0.8cm}, clip=,width=\columnwidth]{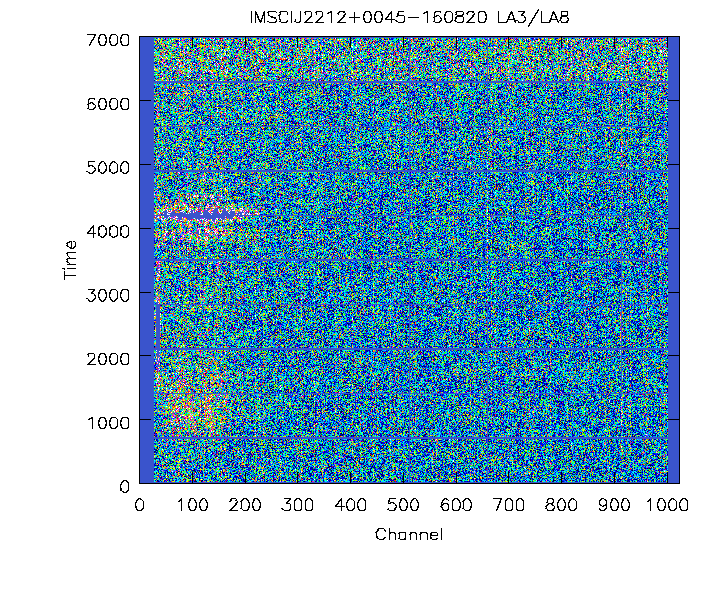}
 \includegraphics[trim={1.5cm, 1.5cm, 0.5cm, 0.8cm}, clip=,width=\columnwidth]{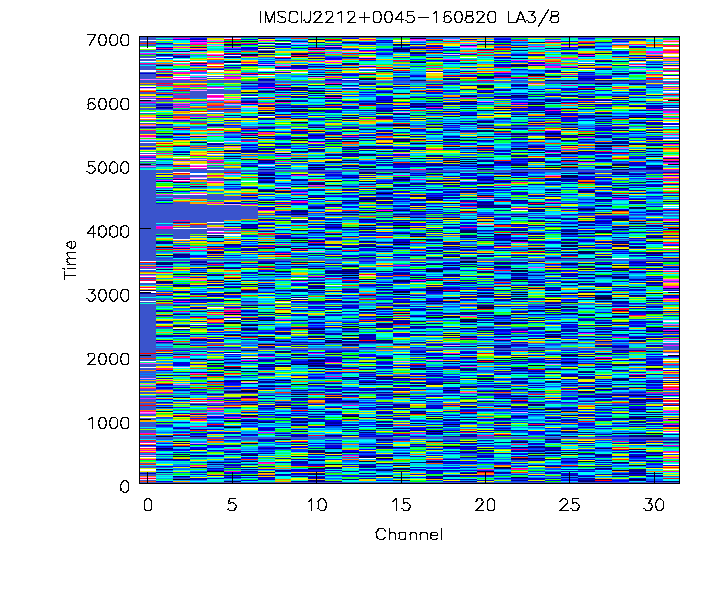}
 \caption{An illustration of RFI removal from the LA observation shown in Fig.~\ref{Fi:RFI}.  The left-hand-side shows the effect of a $5\sigma$ flagging step using the CASA `rflag' algorithm; the bright, narrow-band signals have been removed as well as some of the wider-band signals.  On the right, we have binned in frequency down to 32 channels and made another $5\sigma$ cut, which removes most of the remaining wider-band signals.}
 \label{Fi:RFI_flagging}
\end{figure*}

One of the main motivators for constructing a correlator with such improved spectral resolution was to be able to identify and remove radio-frequency interference (RFI), which is typically confined to narrow frequency bands.  RFI is generally worse at low elevation, due to the location of geostationary satellites.  With the old correlator, this limited the useable declination range, particularly for the SA with its larger primary beam.
\highlight{RFI flagging levels on the SA reached $\approx$\,85\% at $\delta \approx\,2^{\circ}$; in practice observations were rarely made below $\delta=20^{\circ}$.  In Figure~\ref{Fi:RFI} we show part of an observation of a low-declination ($\approx\,0^{\circ}$) field on both the LA and the SA with the new correlator.  The RFI is clearly localized in time- and frequency-space and affected regions of data can be easily excised using automated algorithms such as `rflag' as implemented in CASA, which searches for deviations from the median r.m.s.\ of the data in time and frequency space; see the AIPS cookbook, Section E.5 \footnote{\url{ftp://ftp.aoc.nrao.edu/pub/software/aips/TEXT/PUBL/COOKE.PS.gz}} for more detail on the algorithm.  This removal process is illustrated in Fig.~\ref{Fi:RFI_flagging}.  At $\delta \approx\,0^{\circ}$ the RFI flagging level is now $\approx$\,20\% on the LA and $\approx$\,25 -- 35\% on the SA.  These are comparable to flagging percentages in similar frequency bands at other facilities, for example the Very Large Array which loses $\approx$\,15\% at Ku-band\footnote{see, e.g.\ \url{https://science.nrao.edu/facilities/vla/observing/RFI/jul-2014-d-configuration/Ku-Band_spectra_201407D}}, accounting for the fact that the AMI arrays are more compact and therefore more sensitive to ground-based RFI, and have larger primary beams.}

\subsection{Dynamic range}

\begin{figure*}
 \centering
 \includegraphics[trim={1.5cm 1.5cm 4.7cm 0.8cm},clip=,width=0.4325\linewidth]{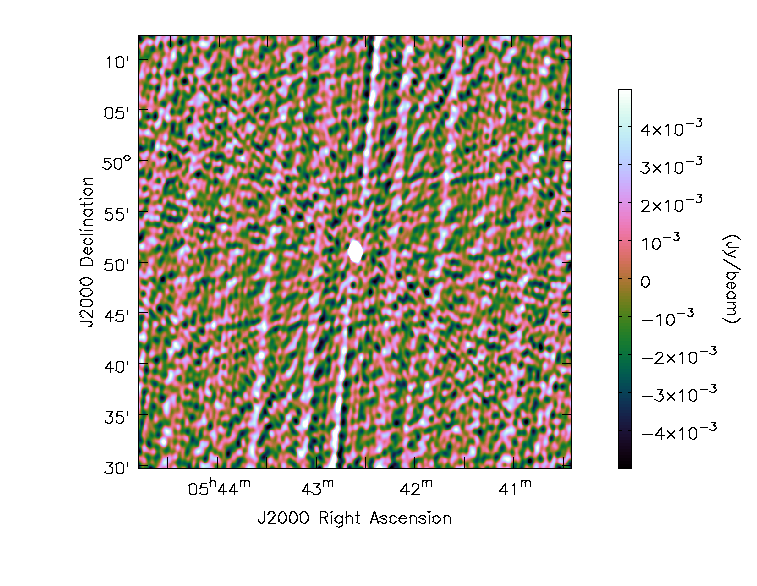}
 \includegraphics[trim={1.5cm 1.5cm 0.6cm 0.8cm},clip=,width=0.5575\linewidth]{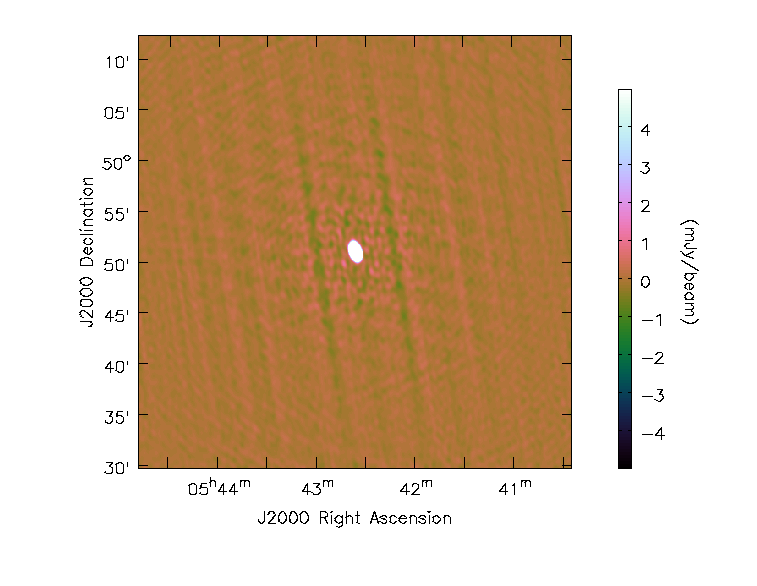}
 \caption{An LA observation of the bright, unresolved source 3C147 using the old (new) correlator on the left (right).  The colour scales are the same, and have been truncated to show the systematic residuals around the source.  These are at most 9.3 (1.1)\,mJy\,beam$^{-1}$, giving a dynamic range of 260 (2300).}
 \label{Fi:3C147_LA}
\end{figure*}

\begin{figure*}
 \centering
 \includegraphics[trim={1.5cm 1.5cm 4.7cm 0.8cm},clip=,width=0.4325\linewidth]{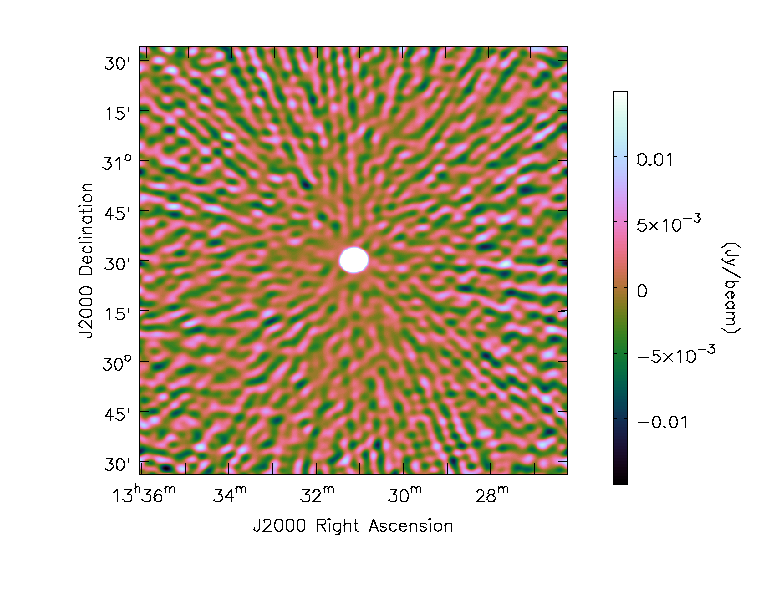}
 \includegraphics[trim={1.5cm 1.5cm 0.6cm 0.8cm},clip=,width=0.5575\linewidth]{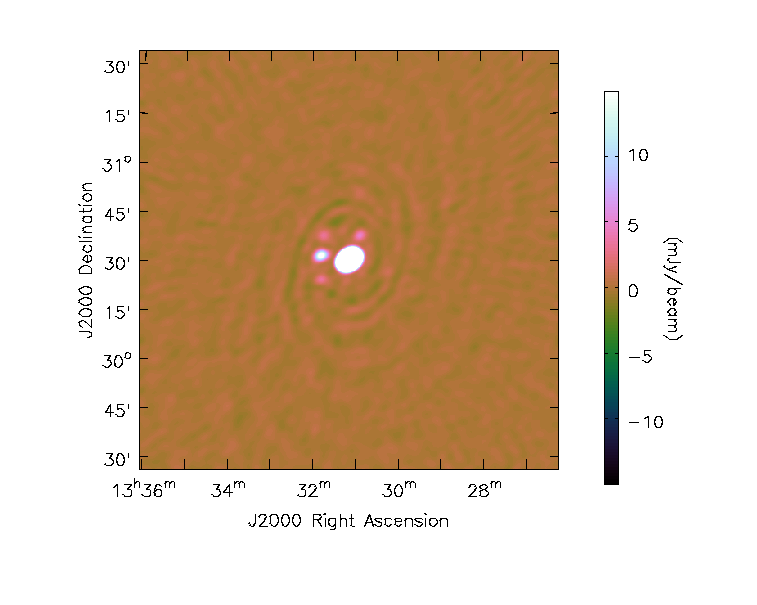}
 \caption{An SA observation of the bright, unresolved source 3C286 using the old (new) correlator on the left (right).  The colour scales are the same, and have been truncated to show the systematic residuals around the source.  These are at most 15 (1.2)\,mJy\,beam$^{-1}$, giving a dynamic range of 230 (3000).  The integration time for the new correlator map was much longer, in order for the thermal noise to fall below the level of the systematic errors.  The triple of sources visible to the east of 3C286 in the new correlator map is real (the central source is FIRST J133148.5+303147); the source to the north of 3C286 is also real (FIRST J133052.9+303807).  They are all under the level of the residuals in the old correlator map.}
 \label{Fi:3C286_SA}
\end{figure*}

The uneven lag spacings of the old correlator limited the dynamic range, since maps of fields near high flux density sources would be contaminated by artefacts introduced by the correlator.  The digital correlator has removed this issue and consequently the dynamic range of the telescope, defined as the ratio of the brightest believable flux to brightest non-believable flux on the map, has increased from $\sim$\,100 to $\sim$\,1000.  Figure~\ref{Fi:3C147_LA} and Figure~\ref{Fi:3C286_SA} show example maps of a bright source produced on the LA and the SA, in comparison to similar maps using old correlator data. The improved dynamic range of the new AMI system is enabling, for example, the observation of galaxy clusters containing bright radio galaxies, which were previously excluded from cluster samples potentially introducing unquantifiable biases.

\subsection{Scientific Collaboration}

AMI has been operating routinely with the upgraded correlator since
January 2016. 
\highlight{The telescope operates under the terms of a Scientific Collaboration Agreement involving the University of Cambridge and the University of Manchester, but welcomes external collaborators. The current scientific programmes include: 
SZ observations of galaxy clusters, for example extending the previous follow-up of \emph{Planck} galaxy clusters \citep{Perrott2015}; 
source surveys, extending the 10C source counts (e.g.\ \citealt{10Ccont}) to lower flux densities and adding to the multi-wavelength legacy datasets available in fields such as Stripe 82 (e.g.\ \citealt{Stripe82} and COSMOS (e.g.\ \citealt{COSMOS});
monitoring of variable and transient radio sources such as supernovae, GRBs, quasars and X-ray binaries;
observations of supernova remnants with reported anomalous microwave emission (AME) detections at low resolution (\citealt{QUIJOTE_W44}, \citealt{Onic_IC443}), to test for the presence of AME at the angular scales measured by AMI.
In each case, the new correlator has allowed AMI to operate more effectively in challenging radio source environments than was previously possible. This is a result of the greater RFI containment -- which allows observations of low-declination fields -- and enhanced dynamic range -- which allows significant improvements in source subtraction capability -- of the new instrument.}

\highlight{The first results using data taken with the upgraded correlator have been published \citep{Munoz-Darias, Mooley2017}, with some work leveraging the new instrument in reobservations of sources originally detected by AMI's analogue correlator \citep{perrott2018}. More results are currently in preparation.}

\section{Conclusions}\label{sec:conclusion}
In this paper we have presented a new digital correlator system for the AMI telescope, which replaces the telescope's previous analogue lag-correlator. This real-time FX correlator is implemented using the popular ``packetised correlator'' architecture, in which interconnect between processors is provided by commercial Ethernet switches. Processing in the new correlator is performed on CASPER open-source hardware; FPGA-based ROACH2 platforms are used to digitise, filter into 1.2\,MHz-wide channels, and correlate a pair of down-converted sub-band signals at 5 Gsps, providing a usable RF band of 13.1--17.9 GHz. This wideband performance has been achieved by clocking the signal processing pipelines on the ROACH2's Xilinx Virtex 6 FPGAs at 312.5\,MHz, requiring optimisation and floorplanning of the firmware design.

The improvement in performance of the new instrument when compared to the original analogue XF correlator is evident; the new correlator achieves over an order of magnitude improvement in imaging dynamic range, and is far more effective at operating in the presence of RFI -- paricularly interference at low-declination observations from geostationary satellites -- owing to superior spectral resolution. As a result of the improved performance, a new scientific collaboration has been established for the operation of AMI, and a broad range of observations are currently being performed with the telescope, which is now highly subscribed. 

\section*{Acknowledgements}
 We acknowledge support from the European Research Council under grant ERC-2012- StG-307215 LODESTONE.
 The images in Figures~\ref{Fi:RFI} to~\ref{Fi:3C286_SA} were prepared using the {\sc Cubehelix} colour palette \citep{Green2011}.
 This work has been supported by the generous donation of FPGA hardware and programming tools by the Xilinx University Program.



\bibliographystyle{mnras}
\bibliography{ami_corr_upgrade_paper}




%


\bsp    
\label{lastpage}
\end{document}